\documentclass[preprint2]{aastex}
\usepackage{listings}
\usepackage{color}
\usepackage{float}
\extrafloats{100}
\usepackage{pifont}
\usepackage{bbding}
\shorttitle{SDSS DR9 Galaxy Clusters from AMF finder}
\shortauthors{Banerjee et al.}

\begin{document}

\title{An Optical Catalog of Galaxy Clusters Obtained
from an Adaptive Matched Filter Finder Applied to SDSS DR9 Data}
\author{P. Banerjee\altaffilmark{1}, T. Szabo\altaffilmark {2}, E. Pierpaoli\altaffilmark{1}, G. Franco\altaffilmark{2},
  M. Ortiz\altaffilmark{2}, A. Oramas\altaffilmark{2}, B. Tornello\altaffilmark{2}} 
 \altaffiltext{1}{Department of Physics and Astronomy, University of Southern
California, Los Angeles, CA 90089, USA} 
\altaffiltext{2}{Department of Physics and Astronomy, Cerritos College, Norwalk, CA 90650, USA}
\email{panchajb@usc.edu,tszabo@cerritos.edu,pierpaol@usc.edu}

\begin{abstract} 
We present a new galaxy cluster catalog constructed from the Sloan Digital Sky Survey Data Release 9 (SDSS DR9) using an Adaptive Matched Filter (AMF) technique.  
Our catalog has 46,479 galaxy
clusters with richness $\Lambda_{200} > 20$ in the redshift range 0.045 $\le z <$ 0.641 in
$\sim$11,500 $deg^{2}$ of the sky. Angular
position, richness, core and virial radii and redshift
estimates for these clusters, as well as their error analysis, are provided as part of this catalog. In addition to the main version of the catalog, we also provide an extended version with a lower richness cut, containing 79,368 clusters. 
 This version, in addition to the clusters in the main catalog, also contains those clusters (with richness $10<\Lambda_{200}<20$) which have a one-to-one match in the DR8 catalog developed by Wen et al (WHL). We obtain probabilities for cluster membership for each galaxy and implement several procedures for the identification and removal of false cluster detections.

We cross-correlate the main AMF DR9 catalog with a number of cluster catalogs in different wavebands (Optical, X--ray).  We compare our catalog with other SDSS-based ones such as the redMaPPer (26,350 clusters) and the Wen et al. (WHL) (132,684 clusters) in the same area of the sky and in the overlapping redshift range. We match 97$\%$ of the richest Abell clusters (Richness group 3), the same as WHL, while redMaPPer matches $\sim 90\%$ of these clusters. Considering AMF DR9 richness bins, redMaPPer does not have one-to-one matches for 70$\%$ of our lowest richness clusters ($20<\Lambda_{200}<40$), while WHL matches 54$\%$ of these missed clusters (not present in redMaPPer). redMaPPer consistently does not possess one-to-one matches for $\sim$20$\%$ AMF DR9 clusters with $\Lambda_{200}>40$, while WHL matches $\geq$70$\%$ of these missed clusters on average. For comparisons with X-ray clusters, we match the AMF catalog with BAX, MCXC and a combined catalog from NORAS and REFLEX. We consistently obtain a greater number of one-to-one matches for X--ray clusters across higher luminosity bins ($L_x>6 \times 10^{44}$ ergs/sec) than redMaPPer while WHL matches the most clusters overall. For the most luminous clusters ($L_x>8$), our catalog performs equivalently to WHL. This new catalog provides a wider sample than redMaPPer while retaining many fewer objects than WHL. 
  \end{abstract}
  
  
\section{Introduction}

Galaxy clusters are the most massive gravitationally bound systems  
in the universe and their study has been pursued for various different 
reasons, both astrophysical and cosmological. Cluster surveys 
allow the determination of clustering properties of density fields
providing information on the 
large scale structure \citep{bahcall88, postman92, coleman92, carlberg96, bahcall97, seljak00, reid10, guo13}.  
Due to their
high galaxy density, implying a higher probability of galaxy collisions, clusters are excellent
laboratories for studying galaxy evolution \citep{dress80,
butch78, garilli99, goto03a, goto03b, mei09, maughan12}.  They
can act as gravitational lenses, providing a way to 
constrain cluster masses, as well as to 
study distant galaxies \citep{blain99,smail02,metcalfe03,
santos04, shel09, oka15}.
The galaxy cluster mass function allows for the determination of  
 several cosmological parameters, including the mass
density, $\Omega_{m}$, and the matter power 
spectrum normalization $\sigma_{8}$ \citep{Pierpa01,reip02,
seljak02,Pierpa03,dahle06,pedersen07,rines07,Rozo09, ang11, ser15} as well as 
the dark energy equation of state \citep{Allen08, man14}
and neutrino masses \citep{Wang05, carb12}. 
Galaxy clusters have also been used to constrain dark matter annihilation \citep{ack10,huang12,ack15}
Cluster abundances and their internal structure have 
been used as tests for modified gravity
\citep{Rapetti08,Diaferio09,jenn12}. 
More recently, the measurement of gravitational redshifts from cluster surveys has provided 
additional opportunities for verifying GR and probing other gravitational physics \citep{wojt11,weg11,kai13,crof13,jim15}.

Galaxy clusters are observed in several bands, such as optical, radio, microwave and X--ray. In the optical, they are observed through the over-density of galaxies 
or their color properties, and in the radio and X--ray 
through the emission of their intra--cluster medium (ICM).
Optical observations were the first ones to be  
performed \citep{abell58} but  X--ray surveys of recent years have discovered  
hundreds of clusters up to high redshifts ($z \simeq 1$) \citep{ebel10,piff11,mehr12,mirk15}.
Radio surveys aiming at detecting galaxy clusters 
through their Sunyaev--Zel'dovich (SZ) effect 
have been undertaken as well \citep{will11,nova12,planck15,bleem15}. 
The X--ray and the optical bands 
trace different physical components, so they represent
very distinct probes of the state of the cluster.
The variety of detection methods and cross-verification between them will provide a better understanding of 
cluster physics and the selection function and biases for each 
detection method, therefore improving the precision 
in deriving cosmological constraints.

Here, we focus on observations of 
clusters in the optical band in the Sloan Digital Sky Survey (SDSS).  
We present a catalog of galaxy clusters constructed from 
an adaptive matched filter (AMF) technique \citep{dong08, 
postman96, kepner99, Kim02} and following up on the work of Szabo et al.
 \citep[hereafter, AMF DR6]{szab11}. SDSS provides data for luminosities in five bands (u-g-r-i-z) and  
redshift estimates for millions of galaxies
in about one-third of the sky. Using this data, it is possible for automated algorithms, such as the AMF, to compile cluster catalogs objectively
by identifying over-densities in galaxy distributions.

We present the catalog extracted from the SDSS
DR9 by 
applying the AMF technique tailored to handle SDSS 
data \citep{dong08}.
We compare this catalog (hereafter AMF DR9) with other optical catalogs such as the Abell catalog \citep{abell58}, the redMaPPer 
 by Rykoff, et al, compiled from SDSS DR8, and with the catalog extracted 
from the SDSS DR8 by \cite{whl11} (hereafter, WHL).  
 Our finder uses
over-densities as a starting point for determining the presence
of a cluster. However, unlike many other cluster determination methods \citep{murph12,ryko14a}, it does not make {\it a priori} assumptions about
the color or number of bright galaxies. Thus, our technique has the potential to 
detect clusters which do not have a 
bright red galaxy. We obtain probability estimates for cluster membership for each galaxy and implement several procedures
to minimize spurious objects and false cluster detections.

In section \ref{data_sect},
we discuss the specifics and nature of the SDSS data used.  We present
the AMF cluster finder and details on its 
application to the SDSS data in section \ref{finder_sect}. In section \ref{finder_sect} we also outline the process used to remove duplicate and spurious clusters. 
Section \ref{cat_sect} discusses the 
characterization of the galaxy cluster catalog.
In section \ref{comp_sect} we compare our catalog with the ones compiled by
WHL, and redMaPPer, as well as with different X-ray cluster catalogs.
Section \ref{conc_sect} is dedicated to 
the conclusions.

A standard $\Lambda$CDM cosmology is assumed,
with $\Omega_{m}$~=~0.3 and $\Omega_{\Lambda}$~=~0.7, and
H$_{0}$~=~100 h km~s$^{-1}$~Mpc$^{-1}$.
 
\section{SDSS Data} \label{data_sect}

The Sloan Digital Sky Survey III \citep{sdss3} follows the SDSS-I and II projects \citep{york00}. SDSS-III uses the 2.5 m wide-field Sloan Foundation Telescope \citep{gunn06} at Apache Point Observatory (APO), and fiber-fed multi-object spectrographs to carry out multiple surveys. Their aim is to study dark energy (Baryon Oscillation Sky Survey, BOSS), to assist in the search for extrasolar planets (MARVELS), and to study the structure of the Milky Way Galaxy(SEGUE-2 and APOGEE) \citep{DR9paper}. 
The SDSS is a five-band CCD 
imaging survey of $14,555$  $\rm{deg}^2$ in 
the North Galactic Cap and the Southern Galactic Cap. The 
imaging survey is carried out in drift-scan mode (opening the camera shutter for long periods to image a continuous strip of the sky). The survey provides data in five SDSS filters, which are labelled u-g-r-i-z and range in wavelength from 355.1 nm to 893.1 nm, and in magnitude limits from 22.0 to 20.5 respectively (with 95 $\%$ completeness for point sources). 
The main galaxy sample has a median redshift of z = 0.1, and there are spectra for luminous red galaxies up to z = 0.7.

In this paper we construct a cluster catalog for Data Release 9 (DR9) of the Sloan Digital Sky Survey and 
have used the photometric redshifts provided by  
Carliles et al. obtained by the Random Forest method. In the AMF DR6 catalog \citep{szab11}, the authors had chosen to use the photometric redshifts 
determined by \citep{oyaizu08} using a Neural Network method. When compared to other catalogs,
a representative sample of preliminary DR9 clusters showed
less scatter in redshift when using Random Forest photometric
redshifts compared to other photometric redshift estimates. 

This ninth data release (DR9) of the SDSS project includes 535,995 new galaxy
spectra (median z$\sim$0.52), 
along with the data presented in previous data releases \citep{DR9paper}. 
The SDSS DR9 fixed a systematic error in the astrometry in the imaging catalogs in DR8 \citep{aiha11}. 

All galaxy measurements were extracted via an SQL query from
the {\tt Galaxy} view on the CasJobs DR9 database
using the {\tt FLAGS} as described in 
Appendix \ref{phot_flags_app}
to obtain a complete sample of galaxies
with good photometry to minimize falsely flagged galaxies. 
The data is then placed in stripes according to the geometry of the survey. 
The current catalog is constructed using $r$-band data information only, it being the band where the 
sensitivity of the SDSS CCDs are the highest, in addition to having a well understood luminosity function
for $m\leq22$ for the r-band. 

\section{The Cluster Finder \label{finder_sect}}

\subsection{Method}

We use an Adaptive Matched Filter cluster finder method \citep{postman96,kawasaki98,kepner99,Kim02 ,dong08,szab11}.
The idea behind the AMF is the matching of the data from the optical galaxy survey (SDSS) with a filter based on a model of
the distribution of galaxies \citep{kepner99}. In our finder, we use the same methodology as that used to construct the AMF DR6 catalog \citep{szab11}. 
The AMF technique does not explicitly use the red sequence to select clusters \citep{Annis02,
Miller05,koester07,ryko14a}. So this technique is not restricted only to old, red 
E/S0 galaxies as in color-based cluster finding methods,  and can theoretically detect clusters 
of any type in color. 

 A cluster radial 
surface density profile, a galaxy luminosity function, and 
redshift information are used to construct filters in 
position, magnitude, and redshift space, from which a cluster likelihood map
is generated. We assume an NFW density profile and a Gaussian model for photometric redshift uncertainties for filters in position and redshift respectively.
For the magnitude filter, we adopt a luminosity profile described
by a central galaxy plus a standard Schechter luminosity function with a characteristic luminosity $L^*$ \citep{schech76}. Cluster and field galaxy luminosity functions are not modeled separately. The existence of brightest cluster galaxies (BCGs) near the
cluster centers is incorporated into the cluster galaxy luminosity model as a component separate from the main Schechter function for satellites.
The BCG luminosity is assumed to follow a single power law with an exponent of 0.22 with cluster richness
and we take the width of the Gaussian to be 0.56 mag. 
We have updated the 
coefficient for the evolution of absolute magnitude of BCGs with respect to cluster richness. An exponent of 0.22 and a gaussian width 
of 0.56 was found to be a better fit to the data, as compared to the previous iteration of the AMF finder \citep{szab11}, with its values of 0.26 and 0.44 respectively.  

The peaks in the likelihood map are where the matches 
between the survey data and the cluster filters are optimized
and thus correspond to candidate cluster centers.  
The algorithm provides the  best-fit estimates of cluster properties 
including redshift, radius and richness. The likelihood of each galaxy belonging to
a particular cluster (and thus a method of membership assessment) is also provided. 

The parameters of the new cluster 
(redshift $z$, core radius $r_c$ and 
richness $\Lambda_{200}$) are varied in order 
to maximize the increase in likelihood. 
The cluster radius $R_{200}$ 
is the distance from the central 
galaxy where the over-density 
of galaxies is 200 times the critical density $\rho_c$. 
We assume that the average galaxy density is representative of 
the mean mass density \citep{jenkins01}.
The richness $\Lambda_{200}$ is the total luminosity within
$R_{200}$ in terms of $L^{*}$, 
where $L^{*}$ brightens with redshift. \citep{Loveday92, Lilly95b,Nagamine01,
 Blanton03,Loveday04, Baldry05,Ilbert05,faber07,bouw15}. The richness is related to the number galaxy counts within a cluster.
  
 The actual center of mass for a galaxy cluster should not necessarily be located on a galaxy, especially 
 given how dark matter comprises most of the mass within a cluster. Therefore, all initially identified clusters with
 $\Lambda_{200} \ge 10$ are re-centered. During the re-centering process,
 we relax the hypothesis that clusters are centered on a galaxy.  The center positions and the 
characterizations of the clusters are refined, and the likelihoods for each cluster are re-evaluated by varying the 
new hypothetical center. Of the original clusters, $\sim32\%$ are found to have their centers displaced by $>$ 10 kpc $h^{-1}$

The presented catalog contains 46,479 clusters with
$\Lambda_{200} \ge$ 20 over an area of $\sim$11,500 deg$^{2}$.

subsection{Catalog Refinements} \label{cat_refine}

	The original cluster catalog, compiled using the methodology mentioned above, contained 54,277 galaxy clusters from the SDSS DR9 data. In the original version of the catalog, there were several instances of clusters lying on the fringes of stripes, or whose footprint showed considerable star interference. There were also instances of one galaxy being assigned to more than one cluster. In this section, we describe the work done to remove these clusters.
	
\subsubsection{Cut-off Clusters}

A `cut-off' cluster is classified as a cluster with an
irregular shape, considerably deviated from circularity in 2-dimensions, possibly cut-off by the edge of a stripe or showing star interference. 
We wrote a cluster cut-off code to 
be sensitive to the shape of the cluster. 
Eventually through the individual inspection of cluster footprints, we established a $16 \%$ deviation from circularity 
as a viable criterion for eliminating a cluster. The finder showed unreliable performance for this difference
from circularity, producing false clusters, or clusters
whose richnesses were unreasonably large. Some allowance
for discrepancies (up to 16$\%$) is necessary due to the granular nature
of the data sample.
This criterion was relaxed slightly with redshift (taking into account the decrease in cluster angular size for clusters and the reduction in visible galaxies due to magnitude limitations at higher z) to 20$\%$ for 
clusters with z $>$ 0.5. There were 2407 clusters which were removed using this method. 
 
 \subsubsection{Probability of Membership Estimates}
 
The finder code itself prevents the detection of the same cluster within a stripe, but we ensure that multiple detections across
different stripes are eliminated.
If a galaxy has a high
probability of being in one cluster, but a low probability to belong to another cluster, it is sorted into the former cluster. 
If a set of galaxies have a substantially high probability of belonging to two clusters, it is probably the same cluster. 
The 48$\%$ cutoff for this 'high' probability was determined by creating a list
of clusters that belonged to adjacent stripes and whose
centers were separated by less than 0.1 degrees and were
within 0.05 of each other in redshift. The galaxy probabilities
for these clusters were analyzed for a trend that would 
provide a threshold for galaxies that belonged to both clusters.
A list of 6264 duplicates were provided, and the lower richness
cluster was slated for removal in each case.

Of the 8671 clusters selected through these criteria, 807 were
selected for removal by both methods. 
There is a possible issue of the shared galaxy criterion
removing a low richness cluster in a match, than the cut-off
criterion removing the higher richness one. The list of clusters to cull
was examined for these cases, and 66 clusters that were
originally removed by this overlap were then restored,
using the lower richness, non-cut-off cluster. 

With the above steps followed, we reduced the AMF DR9 catalog from 54,277 to 46,479 clusters.

\section{The AMF Catalog \label{cat_sect}}

We present the AMF DR9 catalog of galaxy clusters. 

This section presents the main properties 
of the clusters in AMF DR9 such as the richness $\Lambda_{200}$, redshift $z$ and core radius $r_c$, along with the distributions of these quantities for the catalog. The associated error-ranges are also presented. 

We provide a description of the deliverables and the data retrieval procedure in the Appendix.Table \ref{amf_dr9_cat} provides a list of the columns, descriptions and formatting for the catalog.

\subsection{Characteristics}

\begin{figure} [H]
\plotone{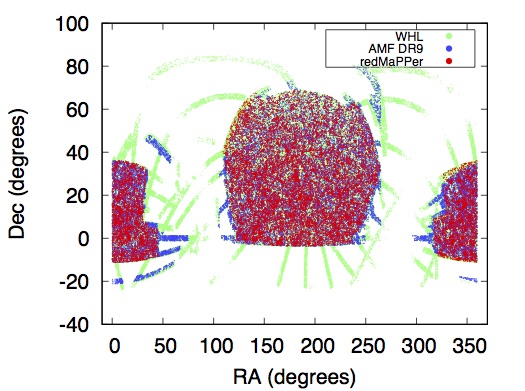}
\caption{Angular Positions of the Clusters found in AMF DR9, redMaPPer and WHL catalogs. In the same area of the sky, 
AMF DR9, redMaPPer and WHL have 46,479, 26,333 and 125,902 clusters respectively}
\label{fig1}
 \end{figure}

The position of the clusters in the sky is represented in fig. \ref{fig1}, for the three major optical catalogs considered in the paper. 

The distribution of cluster density per deg$^{2}$
varies from stripe to stripe, with a variation of 2.7-3.9 for stripes 
in the Northern Galactic Cap and a variation of 3.6-4.9 for stripes
in the Southern Galactic Cap. 
The density variations are most likely due to
large scale structure and the survey depth \citep{peacock01}. 

The minimum $\Lambda_{200}$ of the included clusters is 20, while the maximum richness is 219.36. 
The core radius $r_c$ varies from 0.059 to 1.015 $\times$ $h^{-1}$ Mpc. The redshift range of our catalog is between 0.045 and 0.641.

\begin{figure} [H]
\plotone{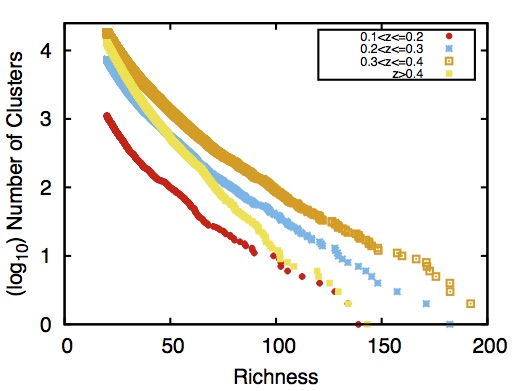}
\caption{Distribution of AMF clusters in richness per redshift bin}
\label{fig2}
 \end{figure}

The distribution of our clusters in richness for different 
redshift bins is shown in Fig. \ref{fig2}.

 The ratio of 
high richness objects over low richness ones decreases 
for increasing redshift, which is reasonable, because the catalog
is likely not complete for $z\sim0.5$. 
Therefore, the drop in the number of very rich clusters
compared to the low richness ones may be due to 
limitations of the finder. 

\begin{figure} [H]
\plotone{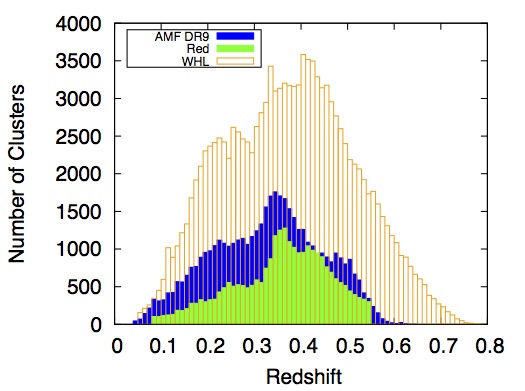}
\caption{Redshift distribution of optical catalogs} 
\label{fig3}
 \end{figure}

Fig \ref{fig3} shows the redshift distribution (in the same region of the sky) of the major optical catalogs used for comparison in this paper (See Section \ref{comp_sect}).

\begin{figure} [H]
\plotone{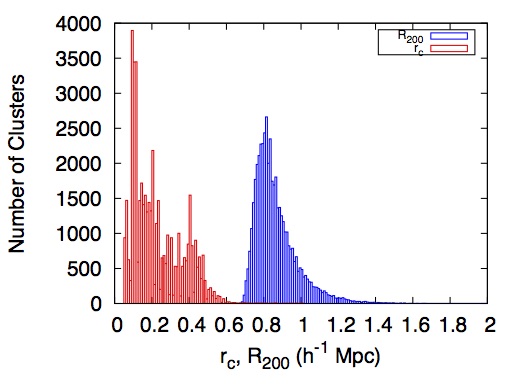}
\caption{The core radius and $R_{200}$ distribution of AMF clusters}
\label{fig4}
 \end{figure}

Histograms of the core radius $r_c$ and R$_{200}$ are presented in Fig. \ref{fig4}. 

 \subsection{Error Determination \label{err_det_sect}}

The AMF finder provides the maximum likelihood values for $\Lambda_{200},z$ and $r_c$ and for the angular position of the cluster center. 
We use the likelihood determination procedure to find errors
in these parameters as follows.
The likelihood for each cluster 
was recomputed on a finer 
grid, considering galaxies within 
2$R_{200}$ and 3$\sigma_{z}$ of the cluster
center. Errors for each quantity are found by
exploring the likelihood surface in two of the parameters, while keeping the
others fixed. The lowest value of error percentages from the various permutations was retained. For example, errors in richness $\Lambda_{200}$ were found by fixing the $r_c$ values and the angular positions of cluster centers while varying $\Lambda_{200}$ and z values. The other likelihood surface variations (varying $r_c$, $\Lambda_{200}$ with cluster center positions and $z$ fixed, and varying 
cluster center positions, $\Lambda_{200}$ while keeping $r_c$ and $z$ fixed) resulted in higher error percentages.

We look for the extrema of the parameters
which give a difference in likelihood from the maximum 
value and then determine the boundaries of the 68$\%$, 90$\%$ and 95$\%$ confidence regions. 
Errors for individual clusters are reported in the catalog,
and include the mean errors for the 68$\%$ and 95$\%$ confidence
intervals for each quantity. 
Using the set of errors generated, we plot the frequency distributions versus one of the defining characteristics of the catalog (z, $\Lambda_{200}$ etc.).

\begin{figure} [H]
\plotone{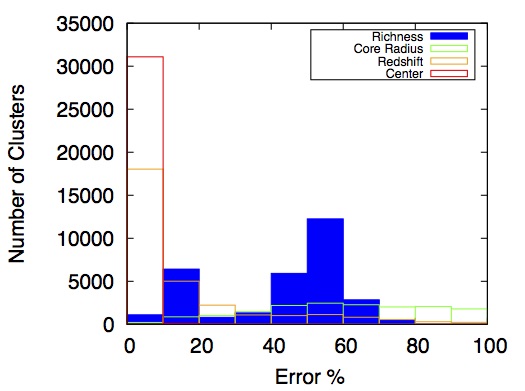}
\caption{Error percentage (68$\%$ confidence range) for $20<\Lambda_{200}<30$} \label{rh2030err_fig}
\label{fig6}
 \end{figure}

\begin{figure} [H]
\plotone{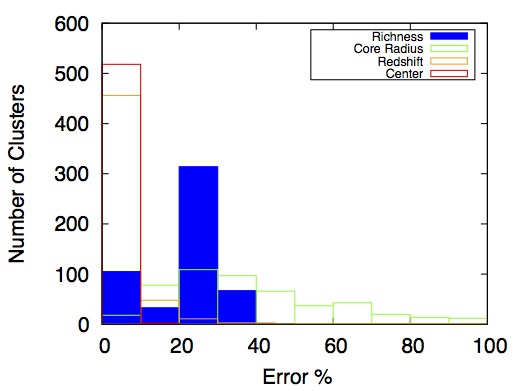}
\caption{Error percentage (68$\%$ confidence range) for $\Lambda_{200}>80$} \label{rhgt80err_fig}
\label{fig7}
 \end{figure}

\begin{figure} [H]
\plotone{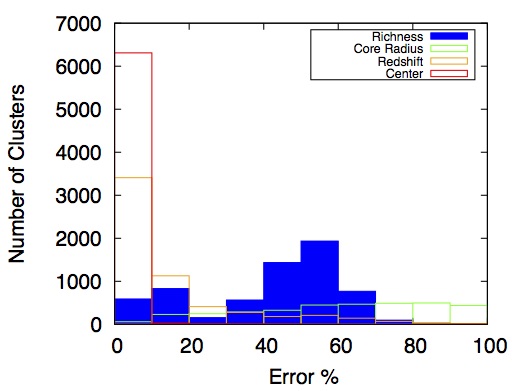}
\caption{Error percentage (68$\%$ confidence range) for $0.1<z<0.2$} \label{zminerr_fig} 
\label{fig8}
 \end{figure}

\begin{figure} [H]
\plotone{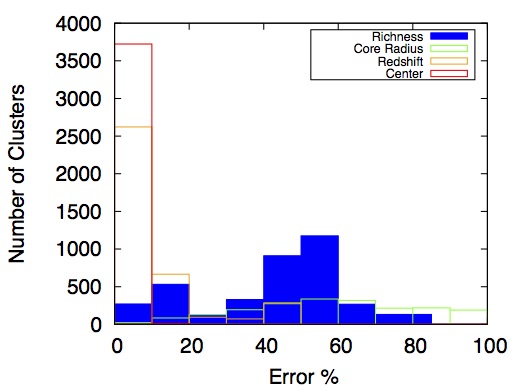}
\caption{Error percentage (68$\%$ confidence range) for $z>0.5$} \label{zmaxerr_fig}
\label{fig9}
 \end{figure}

Figs \ref{rh2030err_fig} - \ref{zmaxerr_fig}  show histograms of the error percentages for the 68$\%$ C.L; for each of the four quantities considered (richness, core radius, redshift and angular position). 
The different plots analyze subsamples of the highest/lowest richness and redshift bins.

The error in richness is seen to
increase as richness decreases, which seems viable as
the addition or subtraction of one galaxy creates a 
larger percentage difference in smaller clusters. 
 Typical errors range from around 50$\%$ for lower richness clusters  
($20 \le  \Lambda_{200} < 30$) to 20-26$\%$ for the highest richness clusters
($\Lambda_{200} >$ 100). Errors in $\Lambda_{200}$ decrease with increasing $r_c$ (Not shown here). 
This is an expected trend as the average $r_c$ for clusters in the published catalog increases
with richness. 

The core radius is found to
be poorly constrained.  Previous studies have indicated that an NFW profile fit for optical clusters
tends to have a large error in $r_c$ \citep{nfw97,cin14,ser16}. We also assume clusters are virialized but the finder is sensitive
to any over-density. 
The error in $r_c$ is smallest
for the richest clusters.  
There is no clear trend in $r_c$ errors for redshift, but the errors
(at a 68$\%$ confidence level) decrease with increasing core radius. 
The high percentage errors in core radius most likely are due to the fact that there are not enough galaxies 
in clusters to accurately map the assumed NFW profile.

The error percentages in redshift are low and decrease steadily as a function of richness. 
There is a slight decreasing trend in the error percentages as we go to higher redshift clusters. 
A similar trend is seen for increasing core-radii. 
The errors in angular positions of the cluster centers are also found to be uniformly low, which implies a stable centering. 

\section{Comparisons with Other Catalogs} \label{comp_sect}

	We check for one-to-one clusters in the AMF catalog with clusters from other catalogs by searching within a given radius and redshift from each AMF cluster center. We adopt a search radius of 1.5 $h^{-1}$ Mpc (which is about twice the minimum value of $R_{200}$ in our catalog) while we allow for a maximum redshift difference of 0.05, which is about twice the typical error in redshift for our catalog. In the rare cases when there is more than one match between AMF DR9 and the other catalog (usually multiple clusters in the other catalog matching one AMF cluster), we retain the matching cluster with the highest optical richness/luminosity. We also take into account the individual cluster errors, in $R_{200}$ and z and determine the number of one-to-one matches in each bin for luminosity or redshift. In addition to this, we check for matches with the redMaPPer and WHL optical catalogs. For comparisons with redMaPPer and WHL, we use the standard matching conditions mentioned for the AMF DR9 catalog.  
	The various labels for the one-to-one matches with our catalog are: 
	\begin{itemize}
	\item AMF DR9: We use the main AMF DR9 catalog for finding one-to-one matches, allowing a search radius of 1.5 $h^{-1}$ Mpc, and a maximum redshift difference of 0.05. AMF DR9 has 44,805 clusters in the redshift range [0.1,0.55]. 
	\item AMF1: In addition to the main catalog, we use clusters with a  minimum $\Lambda_{200}$ of 10 as opposed to 20 for the published catalog. However, we retain only those lower richness clusters that also have one-to-one matches in the WHL catalog under the standard matching conditions (see AMF DR9 above). We allow a maximum redshift difference of 0.05. This extended version of the catalog contains 79,368 clusters overall, and 75,819 clusters in the redshift range [0.1,0.55]. We present this version with the caveat that we have not run the extensive refinement with regards to cluster selection as we have done with the final presented version of the AMF DR9 catalog. 
	\item AMF2: We match with main AMF DR9 catalog but we take three times the individual error in $R_{200}$ of each cluster (that is, we expand the angular radius of the cluster by 3 times the percentage error in $R_{200}$), also using 3 times the redshift error for the same cluster while matching.
		\end{itemize}
		
\subsection{A summary of the catalogs used for comparison}

\subsubsection{Optical Catalogs}

\textit{Abell} The Abell catalog is divided into Richness Groups, ranging from 0-5, in increasing order of number of galaxies within a cluster. In the area considered (see \ref{ab_mat}), there are 1385 Abell clusters in Richness Groups 0-3. 

\textit{redMaPPer} The redMaPPer is a red-sequence cluster finder \citep{ryko14a}. The finder identifies 26350 clusters in $\sim$ 10,000 $deg^2$ of data acquired from SDSS DR8 in the range z $\in [0.08,0.55]$. The richness $\Lambda$ is defined as the sum of the membership probabilities over all galaxies within a scale-radius $R_{\lambda}$. In the redshift range [0.1,0.55], redMaPPer has 26,143 clusters. 

\textit{WHL} The WHL cluster catalog \citep{whl11} uses the photometric redshifts of galaxies from the Sloan Digital Sky Survey III (DR8) to identify 132,684 clusters in the redshift range of z $\in [0.05,0.8]$. The clusters are identified using a friends-of-friends algorithm \citep{botz04}. The cluster richness is defined as $R_{L_*} = L_{200}/L^*$, where $L^*$ is the evolved characteristic luminosity of galaxies in the r-band, and $L_{200}$ is the total r-band luminosity within $R_{200}$. In the redshift range [0.1,0.55], WHL has 116,308 clusters.

\subsubsection{X--ray Catalogs}
 
 We use the following X--ray Catalogs as a basis for comparisons between the different optical catalogs. 
 
 \textit{BAX} We use data from the multi-wavelength X--ray Clusters Database (BAX) \citep{sad04} as updated on February 11, 2011. The BAX database contains 1714 clusters in total in the redshift range 
 z $\in [0.003,5.2]$. In the area considered (see \ref{xray_mat}) and in z $\in$ [0.1,0.55], there are 454 BAX clusters.  
  
 \textit{MCXC} The meta-catalogue of X-ray detected clusters of galaxies (MCXC) \citep{piff11}. The MCXC is based on publicly available ROSAT All Sky Survey-based (NORAS, REFLEX, BCS, SGP, NEP, MACS, and CIZA) and serendipitous (160SD, 400SD, SHARC,WARPS, and EMSS) cluster catalogues. The MCXC database contains 1744 clusters in total in the redshift range z $\in [0.003,1.26]$. In the area considered (see \ref{xray_mat}) and in z $\in$ [0.1,0.55], there are 402 MCXC clusters. 
  
 \textit{NORAS $\&$ REFLEX} The NORAS and REFLEX survey catalogs \citep{bohringer00,bohringer04} describe a statistically complete flux-limited sample (0.1-2.4 keV) of galaxy clusters from the ROSAT-All Sky Survey in the Northern and Southern Galactic caps respectively. We use a combined cluster sample from the 2 catalogs, totaling 825 clusters, of which 150 clusters lie  in the area considered (see \ref{xray_mat}) and in z $\in$ [0.1,0.55].  
 
\subsection{Matches between AMF DR9 and other optical catalogs}

\subsubsection{Matches with Abell clusters} \label{ab_mat}

For matches with the Abell catalog, we bin in Abell Richness Groups and implement an areacut which overlaps the area coverages of the AMF DR9, redMaPPer and WHL catalogs. 
We divide the footprint into three segments and check for one-to-one matches between the aforementioned optical catalogs and the Abell clusters in the following RA and Dec ranges:
\begin{enumerate}
\item RA $\in$ (0,30), Dec $\in$ (-10,30)
\item RA $\in$ (130,245), Dec $\in$ (-3,70)
\item RA $\in$ (320,36), Dec $\in$ (-5,30)
\end{enumerate}

Table \ref{abell_table} show the relative numbers (matching fractions) of one-to-one matches between the Abell catalog and the other optical cluster catalogs, for various Abell Richness Groups.  

We check for one-to-one cluster matches with the Abell cluster catalog, with a $97\%$ matching rate 
for the richest Abell clusters (Richness Group 3). The Abell clusters, especially the richest ones, are well-established galaxy clusters and it is reasonable that the AMF DR9 has a high rate of one-to-one matches with these clusters. These are the brightest clusters in the sky, visually identifiable without the help of algorithms, so even with the caveat of not possessing redshift estimates for these clusters, we expect most of these clusters to be identified in any optical cluster catalog. AMF DR9 finds as many one-to-one Abell matches for Richness Group 3 as WHL, while consistently matching a higher cluster fraction than redMaPPer across all Abell Richness Groups. For this comparison, the number of clusters in AMF, redMaPPer and WHL are 44,805, 26,143, and 116,308 respectively. 

\begin{deluxetable}{llccccc} 
\tabletypesize{\footnotesize}
\tablecolumns{5}
\tablewidth{0pt}
\tablecaption{Abell Catalog Matches \label{abell_table}}
\tablehead{
\multicolumn{1}{c}{Richness} & \multicolumn{1}{c}{Clusters} & \multicolumn{1}{c}{AMF DR9} 
& \multicolumn{1}{c}{redMaPPer} & \multicolumn{1}{c}{WHL} 
}
\startdata
0 & 410 & 0.68 & 0.55 & 0.86\\
1 & 742 & 0.81 & 0.74 & 0.91\\
2 & 195 & 0.92 & 0.84 & 0.95\\
3 & 38  & 0.97 & 0.90 & 0.97\\
\enddata
\tablecomments{Fraction of one-to-one matches with Abell clusters (without redshift estimates) for different richness bins. Richness here refers to the Abell Richness Group. Column 1 lists the Abell Richness Group, Column 2 lists number of clusters in that Group, the remaining columns indicate the fraction of total Abell clusters matched in that richness bin by AMF DR9, redMaPPer and WHL respectively.}
\end{deluxetable}

\subsubsection{Matches with redMaPPer and  WHL}

We examine, in the same area coverage, and in the same overlapping redshift range z $\in$ [0.1,0.55], how many AMF clusters (in richness bins) do not posses one-to-one matches in redMaPPer and whether these missed AMF clusters have one-on-one matches in WHL under the same matching conditions. Table \ref{red_comprich_table} shows these comparisons. Table \ref{red_compz_table} displays the same comparison but for AMF DR9 redshift bins. From these tables it is evident that among the clusters present in the AMF DR9 catalog (with no one-to-one matches with redMaPPer), a large fraction ($> 70\%$ for AMF clusters with $\Lambda_{200} > 60$) is matched by the other optical catalog considered (WHL). In the redshift range z $\in$ [0.3,0.5] redMaPPer misses $\sim 40\%$ of the AMF DR9 clusters considered, while, of these clusters without a one-to-one match in redMaPPer, WHL matches $\sim 50\%$.

\begin{deluxetable}{llcccccc}
\tabletypesize{\footnotesize}
\tablewidth{0pt}
\tablecolumns{7}
\tablecaption{Relative Matching - AMF DR9 vs redMaPPer (AMF Richness Bins) \label{red_comprich_table}}
\tablehead{
\multicolumn{1}{c}{} & \multicolumn{1}{c}{} &
\multicolumn{2}{c}{$$} & 
\\
\multicolumn{1}{c}{$(\Lambda_{200})_{min}$} &
\multicolumn{1}{c}{$(\Lambda_{200})_{max}$} &
\multicolumn{1}{c}{AMF DR9 Clusters} & \multicolumn{1}{c}{redMaPPer Misses} &
\multicolumn{1}{c}{WHL Matches} 
}
\startdata
20 & 40 & 39004 & 0.70  & 0.54 \\
40 & 60 & 4431 & 0.26 & 0.72 \\
60 & 80 & 920 & 0.21 & 0.77  \\
80 & 100 & 282 & 0.21 & 0.69 \\
100 & \nodata & 168 & 0.20 & 0.71 \\
\enddata
\tablecomments{We list the fraction of AMF DR9 clusters not possessing one-to-one matches with redMaPPer in each AMF $\Lambda_{200}$ bin. Then we check for one-to-one matches for these AMF clusters (not occurring in redMaPPer) with WHL and list these fractions. For example, in the range $40<\Lambda_{200}<60$, AMF DR9 has 4431 clusters, out of which redMaPPer matches 3268 clusters, thus missing 1163 clusters (or 0.26 of the AMF DR9 clusters). Among these 1163 clusters, 840 are found to have one-to-one matches with WHL, under the same matching conditions (A matching fraction of 0.72). All catalogs are matched in the overlapping redshift range z $\in$ [0.1,0.55].}
\end{deluxetable}

\begin{deluxetable}{llcccccc}
\tabletypesize{\footnotesize}
\tablewidth{0pt}
\tablecolumns{7}
\tablecaption{Relative Matching - AMF DR9 vs redMaPPer (AMF z bins) \label{red_compz_table}}
\tablehead{
\multicolumn{1}{c}{} & \multicolumn{1}{c}{} &
\multicolumn{2}{c}{$$} & \\
\multicolumn{1}{c}{$z_{min}$} &
\multicolumn{1}{c}{$z_{max}$} &
\multicolumn{1}{c}{AMF DR9 Clusters} & \multicolumn{1}{c}{redMaPPer Misses} &
\multicolumn{1}{c}{WHL Matches} 
}
\startdata
0.1 & 0.2 & 6325 & 0.72 & 0.68  \\
0.2 & 0.3 & 10831 & 0.66 & 0.64  \\
0.3 & 0.4 & 15179 & 0.60 & 0.50  \\
0.4 & 0.5 & 9597 & 0.59 & 0.47   \\
0.5 & 0.55 & 2873 & 0.71 & 0.37 \\
\enddata
\tablecomments{We list the number of AMF DR9 clusters not possessing one-to-one matches with redMaPPer in each AMF $z$ bin. Then we check for one-to-one matches for these AMF clusters (not occurring in redMaPPer) with WHL and list these fractions. For example, in the range $0.2<z<0.3$, AMF DR9 has 10831 clusters, out of which redMaPPer matches 3685 clusters, thus missing 7146 clusters (or 0.66 of the AMF DR9 clusters). Among these 7146 clusters, 4601 are found to have one-to-one matches with WHL, under the same matching conditions (A matching fraction of 0.64). All catalogs are matched in the overlapping redshift range z $\in$ [0.1,0.55].}
\end{deluxetable}

\subsection{Matches with X--ray catalogs} \label{xray_mat}

\begin{figure}[H]
\plotone{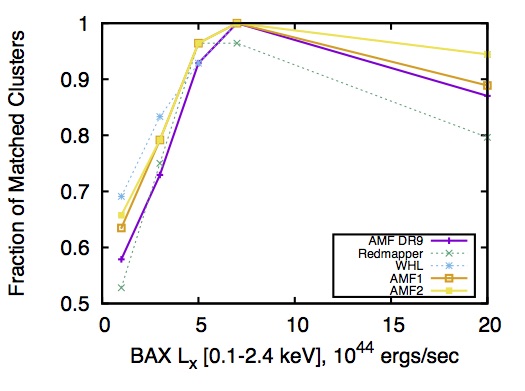} 
\caption{Fraction of BAX clusters matched as a function of X--ray Luminosity. The percentages are computed with respect to each bin in luminosity i.e. number of clusters matched by the optical catalog/total clusters in that bin. The bins for increasing luminosity in BAX have 178, 48, 28, 28 and 54 clusters respectively} \label{bax_lx}
\label{fig12}
\end{figure}

\begin{figure}[H]
\plotone{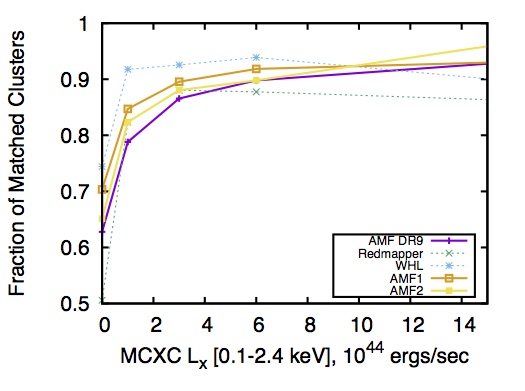} \caption{Fraction of MCXC clusters matched as a function of X--ray luminosity. The percentages are computed with respect to each bin in luminosity i.e. clusters matched by the optical catalog/total clusters in that bin. The bins for increasing luminosity in MCXC have 172, 85, 67, 49 and 29 clusters respectively.} \label{mcxc_lx}
\label{fig13}
\end{figure}

\begin{figure}[H]
\plotone{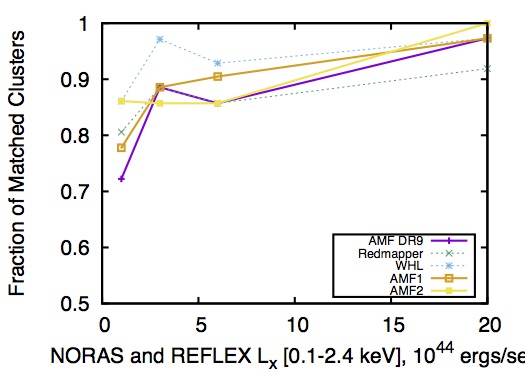} \caption{Fraction of NORAS and REFLEX clusters matched as a function of X--ray luminosity. The percentages are computed with respect to each bin in luminosity i.e. clusters matched by the optical catalog/total clusters in that bin. The bins for increasing luminosity in NORAS and REFLEX have 36, 35, 42 and 37 clusters respectively.} \label{refnor_lx}
\label{fig14}
\end{figure}

The AMF DR9 and other optical catalogs are compared to various X--ray catalogs (BAX, MCXC and a combination of the NORAS and REFLEX catalogs) in order to test the luminosity completeness of the optical catalogs. For matches, we bin in X--ray luminosity and implement an areacut which overlaps the area coverages of the AMF DR9, redMaPPer and WHL catalogs. We divide the footprint into three segments and check for one-to-one matches between the aforementioned optical catalogs and the X--ray clusters in the following RA and Dec ranges:
\begin{enumerate}
\item RA $\in$ (0,30), Dec $\in$ (-10,35)
\item RA $\in$ (120,245), Dec $\in$ (-3,60)
\item RA $\in$ (320,36), Dec $\in$ (-10,30)
\end{enumerate}
We compare the catalogs within the overlapping redshift z $\in$ [0.1,0.55]. We ensure that each bin considered has at least 20 clusters, to minimize the chance of statistical anomalies. Tables \ref{baxlx_comp_table}, \ref{mcxclx_comp_table}  and \ref{refnorlx_comp_table} provide the matching fractions for X--ray luminosity bins with the various optical catalogs considered. The completeness plots for luminosity are shown in Figs \ref{bax_lx} - \ref{refnor_lx}. 

We consistently obtain a greater number of one-to-one matches for X--ray clusters across higher luminosity bins ($L_x>6$) than redMaPPer while WHL matches the most clusters overall. For the most luminous clusters ($L_x>8$), our catalog performs equivalently to WHL. WHL identifies fainter objects as clusters (it has a lower richness cut), and has many more clusters as a result. redMaPPer also contains fewer clusters per unit volume than the other optical catalogs considered \citep{ryko14b}, so it is to be expected that WHL seems more 'complete' for X--ray clusters. WHL also has a lower richness cut ($R_{L_*} > 12$) for clusters so it matches a higher number of lower luminosity X--ray clusters compared to AMF DR9 and redMaPPer. When we match with the extended AMF catalog (AMF1), we see a marked improvement in number of X--ray cluster matches for $L_x<4$ X--ray clusters, while implementing a matching with $R_{200}$ and z errors (AMF2) results in more matches for higher luminosity X--ray clusters ($L_x>4$). The matching fractions for  X--ray clusters  with $L_x>6$ are more than or equal to that from WHL. For these comparisons, the number of clusters in AMF, redMaPPer and WHL are 44,805, 26,143 and 116,308 respectively. 

An investigation of the high luminosity X--ray clusters does not reveal a clear trend in clusters without a one-to-one optical match. Investigating the X--ray cluster center coordinates on the SDSS DR9 Skyserver, we usually find a bright object (typically a star with r-band mag $\sim$ 10) within 2 arc minutes. In some cases, we match the missed X--ray cluster with the AMF DR9  catalog if we relax the original redshift matching criterion ($|dz|<=0.05$). This leads us to believe that our finder might be triggered by very bright objects nearby, identifying a cluster with the same galactic coordinates but with a lower value of z. The reasons why other optical catalogs (redMaPPer and WHL) do not match these luminous X--ray clusters merit further investigation. We test the reasonable assertion that the increased number of one-to-one matches after including the extra clusters for AMF1 are just random variations. The extended AMF catalog adds 32,889 clusters (with a corresponding WHL cluster) to the main AMF DR9 catalog, so we generate 10 random catalogs with 32,889 clusters each (with the RA and Dec randomized) with the same redshift as the added clusters. We then compare the numbers of one-to-one matches between these randomly generated catalogs and the actual added clusters. We find that the randomly generated catalogs uniformly match less X--ray clusters than the added clusters to the extended catalog. On average, the randomly generated clusters match $< 30\%$ of the extra clusters for X--ray clusters with $L_x<4$, and $<50\%$ of the extra AMF clusters for X--ray clusters with $L_x>6$.  

\begin{deluxetable}{llcccccccc}
\tabletypesize{\footnotesize}
\tablewidth{0pt}
\tablecolumns{9}
\tablecaption{Matches with Optical Catalogs (BAX $L_x$ bins) \label{baxlx_comp_table}}
\tablehead{
\multicolumn{1}{c}{} & \multicolumn{1}{c}{} &
\multicolumn{1}{c}{$$} &
\multicolumn{1}{c}{} \\
\multicolumn{1}{c}{$(L_x)_{min}$} &
\multicolumn{1}{c}{$(L_x)_{max}$} &
\multicolumn{1}{c}{BAX Clusters} & \multicolumn{1}{c}{AMF DR9} &
\multicolumn{1}{c}{redMaPPer} & \multicolumn{1}{c}{WHL} &
\multicolumn{1}{c}{AMF1} & \multicolumn{1}{c}{AMF2} 
}
\startdata
\nodata & 2 & 178 & 0.58 & 0.53 & 0.69  & 0.64  & 0.65\\
2 & 4 & 48 & 0.73 & 0.75 & 0.83 & 0.79 & 0.79\\
4 & 6 & 28 & 0.93 & 0.96 & 0.93  & 0.96 & 0.96\\
6 & 8 & 28 & 1.0 & 0.96 & 1.0 & 1.0 & 1.0\\
8 & \nodata & 54 & 0.87 & 0.80 & 0.94 & 0.89 & 0.94\\
\enddata
\tablecomments{Matching fractions for BAX luminosity bins with various optical catalogs. The table lists the number of BAX clusters present in the various luminosity bins, and the subsequent fraction of matches in the AMF DR9, redMaPPer and WHL catalogs. The AMF1 column lists the matching fractions with the extended catalog. The AMF2 column lists the matching fractions with the main AMF DR9 catalog but with errors in $R_{200}$ and z taken into account.}
\end{deluxetable} 

\begin{deluxetable}{llcccccccc}
\tabletypesize{\footnotesize}
\tablewidth{0pt}
\tablecolumns{9}
\tablecaption{Matches with Optical Catalogs (MCXC $L_x$ bins) \label{mcxclx_comp_table}}
\tablehead{
\multicolumn{1}{c}{} & \multicolumn{1}{c}{} &
\multicolumn{1}{c}{$$} &
\multicolumn{1}{c}{} \\
\multicolumn{1}{c}{$(L_x)_{min}$} &
\multicolumn{1}{c}{$(L_x)_{max}$} &
\multicolumn{1}{c}{MCXC Clusters} & \multicolumn{1}{c}{AMF DR9} &
\multicolumn{1}{c}{redMaPPer} & \multicolumn{1}{c}{WHL} &
\multicolumn{1}{c}{AMF1} & \multicolumn{1}{c}{AMF2}
}
\startdata
\nodata & 1 & 172 & 0.63 & 0.51 & 0.74 & 0.70 & 0.65\\
1 & 2 & 85 & 0.79 & 0.82 & 0.92 & 0.85 & 0.82\\
2 & 4 & 67 & 0.87 & 0.88 & 0.93 & 0.90 & 0.88 \\
4 & 8 & 49 & 0.90 & 0.88 & 0.94 & 0.92 & 0.90 \\
8 & \nodata & 29 & 0.93 & 0.86 & 0.90 & 0.93 & 0.97\\
\enddata
\tablecomments{Matching fractions for MCXC luminosity bins with the various optical catalogs. The table lists the number of MCXC clusters present in the various luminosity bins, and the subsequent fraction of matches in the AMF DR9, redMaPPer and WHL catalogs. The AMF1 column lists the matching fractions with the extended catalog. The AMF2 column lists the matching fractions with the main AMF DR9 catalog but with errors in $R_{200}$ and z taken into account.}
\end{deluxetable} 

\begin{deluxetable}{llcccccccc}
\tabletypesize{\footnotesize}
\tablewidth{0pt}
\tablecolumns{9}
\tablecaption{Matches with Optical Catalogs (NORAS and REFLEX $L_x$ bins) \label{refnorlx_comp_table}}
\tablehead{
\multicolumn{1}{c}{} & \multicolumn{1}{c}{} &
\multicolumn{1}{c}{$$} &
\multicolumn{1}{c}{} \\
\multicolumn{1}{c}{$(L_x)_{min}$} &
\multicolumn{1}{c}{$(L_x)_{max}$} &
\multicolumn{1}{c}{X-ray Clusters} & \multicolumn{1}{c}{AMF DR9} &
\multicolumn{1}{c}{redMaPPer} & \multicolumn{1}{c}{WHL} &
\multicolumn{1}{c}{AMF1} & \multicolumn{1}{c}{AMF2} 
}
\startdata
\nodata & 2 & 36 & 0.72 & 0.81 & 0.86 & 0.78 & 0.86 \\
2 & 4 & 35 & 0.89 & 089 & 0.97 & 0.89 & 0.86 \\
4 & 8 & 42 & 0.86 & 0.86 & 0.93 & 0.91 & 0.86 \\
8 & \nodata & 37 & 0.97 & 0.92 & 0.97 & 0.97 & 1.0 \\
\enddata
\tablecomments{Matching fractions for NORAS and REFLEX luminosity bins with the various optical catalogs. The table lists the number of NORAS and REFLEX clusters present in the various luminosity bins, and the subsequent fraction of matches in the AMF DR9, redMaPPer and WHL catalogs. The AMF1 column lists the matching fractions with the extended catalog. The AMF2 column lists the matching fractions with the main AMF DR9 catalog but with errors in $R_{200}$ and z taken into account.}
\end{deluxetable} 

\section{Conclusions \label{conc_sect}}

In this paper we present a new optical catalog of 46,479 galaxy clusters
with richness $\Lambda_{200}>20$ extracted from the SDSS DR9. The catalog extends from 
z=0.045 to 0.641 on an area of 11,500 deg$^2$.
The catalog was constructed using a maximum likelihood 
technique based on a matched filter approach. This   
allows for the simultaneous determination of richness $\Lambda_{200}$,  
core radius $r_c$, redshift, with associated errors. 
The technique does not rely on the red sequence for cluster detection, 
potentially allowing for the detection of galaxy clusters that 
do not possess a central luminous red galaxy. 
 In addition to the main version of the catalog (AMF DR9), we also provide a second extended version with a lower richness cut. 
 This version, in addition to the main catalog, also contains those clusters (with richness $10<\Lambda_{200}<20$) which have a one-to-one match in the DR8 catalog developed by Wen et al (WHL).
  This extended version of the catalog contains 79,368 clusters. Details for retrieval of the catalog data are provided in Appendix \ref{catalog}.

We cross-correlate the main AMF DR9 catalog with a number of cluster catalogs in different wavebands (Optical, X--ray). We compare our catalog with other SDSS-based ones such as the redMaPPer catalog (26,350 clusters) and the Wen et al. (WHL) catalog (132,684 clusters) in the same area of the sky and in the overlapping redshift range. We match 97$\%$ of the richest Abell clusters (Richness group 3), the same as WHL, while redMaPPer matches $\sim 90\%$ of these clusters. Considering AMF DR9 richness bins, redMaPPer does not have one-to-one matches for 70$\%$ of our lowest richness clusters ($20<\Lambda_{200}<40$), while WHL matches 54$\%$ of these missed clusters (not present in redMaPPer). redMaPPer consistently does not possess one-to-one matches for $\sim$20$\%$ AMF DR9 clusters with $\Lambda_{200}>40$, while WHL matches $\geq$70$\%$ of these missed clusters on average. In the redshift range z $\in$ [0.3,0.5] redMaPPer misses $\sim 40\%$ of the AMF DR9 clusters considered, while, of these clusters without a one-to-one match in redMaPPer, WHL matches $\sim 50\%$. 

 Optical catalogs differ in their data selection methodologies, construction of photometric redshift estimates as well as the cluster identification criteria, but we compare AMF DR9, redMaPPer and WHL across overlapping redshift ranges in the same area of the sky. For the comparisons, the AMF DR9, redMaPPer and WHL have 26,143, 44,805 and 116,308 clusters respectively. We consistently obtain a greater number of one-to-one matches for X--ray clusters across higher luminosity bins ($L_x>6$) than redMaPPer while WHL matches the most clusters overall. For the most luminous clusters ($L_x>8$), our catalog performs equivalently to WHL. When we match with the extended AMF catalog, we see a marked improvement in number of X--ray cluster matches for $L_x<4$ X--ray clusters, while implementing a matching with $R_{200}$ and z errors results in more matches for higher luminosity X--ray clusters ($L_x>4$). The matching fractions for $L_x>6$ X--ray clusters are more than or equal to that from WHL.

The AMF DR9 is a new optical cluster catalog with multiple checks in place across several stages, from data selection to cluster finding, 
to ensure no objects are falsely flagged as clusters and with fairly stringent criteria for identifying an object as a cluster, thus ensuring the reliability of the parameters provided. 

\section*{Acknowledgments}

T.S., G.F., A.O., M.O., and B.T. acknowledge support from 
Department Of Education grant P031C110164-15. 
EP thanks the Aspen Center for Physics for their hospitality during the preparation of this work. 
PB was partially supported by the WiSE major support for faculty, awarded to EP.
Funding for the Sloan Digital Sky Survey (SDSS) has been
provided by the Alfred P. Sloan Foundation, the Participating Institutions, the National Aeronautics and Space Administration, the National Science Foundation, the U.S. Department of Energy, the Japanese Monbukagakusho, and the Max Planck Society. The SDSS Web site is http://www.sdss.org/. The SDSS is managed by the Astrophysical Research Consortium (ARC) for the Participating Institutions.

\appendix
\section{Appendix: Photometric Flags for the SDSS Galaxy Sample
\label{phot_flags_app}}
The following flags were set in our CAS Jobs queries
to the SDSS database in order to ensure that our 
data sample had good photometry. 

\begin{lstlisting}
 select g.objid as id_no, g.ra as ra, g.dec as dec, isnull(z.z,-9999) as z,
  isnull(z.zerr,-9999) as zerr, 
   isnull(zrf.z,-9999) as zrf, isnull(zrf.zerr,-9999) as zrferr,
  isnull(s.z,-9999) as sz, isnull(s.zerr,-9999) as szerr, 
  isnull(s.scienceprimary,-99) as sci_prim,
  g.dered_u as mu, g.dered_g as mg, g.dered_r as mr, 
  g.dered_i as mi, g.dered_z as mz,
  g.err_u as muerr, g.err_g as mgerr, g.err_r as mrerr, 
  g.err_i as mierr, g.err_z as mzerr
into mydb.ra150_200_dec_35_45_jul2014
from galaxy g
  left outer join photoz as z on z.objid=g.objid
  left outer join photozrf as zrf on zrf.objid=g.objid
  left outer join specobj as s on s.bestobjid=g.objid
where ((case when (g.type_g=3) then 1 else 0 end) + 
(case when (g.type_r=3) then 1 else 0 end) + 
(case when (g.type_i=3) then 1 else 0 end)) > 1
AND ((case when (g.dered_g < 11) then 1 else 0 end) 
+ (case when (g.dered_r < 11) then 1 else 0 end) 
+ (case when (g.dered_i < 11) then 1 else 0 end)) < 1
  AND g.ra >= 150.0 AND g.ra < 200.0
  AND g.dec >= 35.0 AND g.dec < 45.0
 AND ((g.flags_r & 0x10000000) != 0)
  AND ((g.flags_r & 0x8100000c00a0) = 0)
  AND (((g.flags_r & 0x400000000000) = 0) or (g.psfmagerr_r$\leq$0.2))
AND (((g.flags_r & 0x100000000000) = 0) or (g.flags_r & 0x1000) = 0)
AND abs(g.dered_g - g.dered_r) < 4.5
AND abs(g.dered_r - g.dered_i) < 4.5
  AND g.dered_r<=22.0
 \end{lstlisting}

 From the SDSS DR9 database we select objects of type Galaxy, with clean photometry, using the 
 flags recommended by SDSS. We also specify that the object cannot be saturated, or have its center 
 very close to a saturated pixel. The apparent magnitude in the r-band is $<$ 22. 	
 Further exclusions are for objects 
with an apparent (extinction-corrected) magnitude $ <$ 11 (no galaxy
at z=0.05 or more distant appears this bright in the sky), and for
objects that have color profiles that do not match a galaxy.
There should not be a difference of 4.5 magnitudes (a factor
of 63 in brightness) between adjacent bands for any population
of stars. 

\section{Appendix: Retrieving The Catalog} \label{catalog}

Table \ref{amf_dr9_cat} lists the columns in the main AMF DR9 catalog and the extended catalog, 

  \begin{deluxetable}{llccc}
\tabletypesize{\footnotesize}
\tablewidth{0pt}
\tablecolumns{3}
\tablecaption{AMF DR9 Cluster Catalog Columns \label{amf_dr9_cat}}
\tablehead{
\multicolumn{1}{c}{Column Number} &
\multicolumn{1}{c}{Format} &
\multicolumn{1}{c}{Description} 
}
\startdata
1 & I5 & AMF DR9 cluster number\\
2 &	F10.4 & Right Ascension of the cluster\\
3 & F10.4 & Declination of the cluster\\
4 & F8.4 & Redshift estimate\\
5 & F10.4 & Likelihood\\
6 & F10.4 & Richness ($\Lambda_{200}$)\\
7 & F7.3 & $R_{200}$ \\
8 & F7.3 & Core radius for the cluster in $h^{-1}$ Mpc\\
9 & F8.3 & Concentration, $R_{200}/r_c$\\
10 & F9.4 & Stripe.Richness\\ 
11 & I5 & BAX Matching Cluster ID (if present)\\
12 & I5 & MCXC Matching Cluster ID (if present)\\
13 & I5 & redMaPPer Matching Cluster ID (if present)\\
14 & I5 & WHL Matching Cluster ID (if present)\\
\enddata
\tablecomments{Stripe.Richness gives the stripe in which the cluster is located, followed by the ranking of the 
cluster in richness within that stripe}
\end{deluxetable}

In addition to these, we also provide a table of the error estimates for $\Lambda_{200}$, $r_c$ , z and
$R_{200}$ for each cluster for the catalog, where available. These columns are listed below in Table \ref{amf_err_cat}. 

 \begin{deluxetable}{llccc}
\tabletypesize{\footnotesize}
\tablewidth{0pt}
\tablecolumns{3}
\tablecaption{AMF DR9 Cluster Catalog Error Ranges \label{amf_err_cat}}
\tablehead{
\multicolumn{1}{c}{Column Number} &
\multicolumn{1}{c}{Format} &
\multicolumn{1}{c}{Description} 
}
\startdata
1 & I5 & AMF DR9 cluster number\\
2 & F10.4 & Richness ($\Lambda_{200}$) error (68 $\%$) \\
3 & F10.4 & $\Lambda_{200}$ error (95 $\%$)\\
4 & F10.4 & Core radius ($r_c$) error (68 $\%$) \\
5 & F10.4 & $r_c$ error (95 $\%$)\\
6 & F10.4 & Redshift ($z$) error (68$\%$) \\
7 & F10.3 & $z$ error (95$\%$) \\
8 & F10.3 & Error in centering (68 $\%$) \\
9 & F10.3 & Error in centering (95 $\%$)
\enddata
\tablecomments{The table lists the mean errors in each of the quantities mentioned (in the respective confidence intervals)}
\end{deluxetable}

We also provide a list of probabilities for each galaxy belonging to a cluster. All data are provided in the ASCII format. The catalogs will be uploaded to a website once the paper is accepted for publication.



\end{document}